\documentclass[%
 aip,
 amsmath,amssymb,
 reprint,%
]{revtex4-1}

\usepackage{graphicx}
\usepackage{dcolumn}
\usepackage{bm}

\usepackage[utf8]{inputenc}
\usepackage[T1]{fontenc}
\usepackage{mathptmx}

\begin{document}

\preprint{AIP/123-QED}

\title{Electro-optic eigenfrequency tuning of potassium tantalate-niobate microresonators}

\author{Jan Szabados}
\email{jan.szabados@imtek.uni-freiburg.de.}
\author{Christoph S. Werner}
\author{Simon J. Herr}
\affiliation{ 
	Laboratory for Optical Systems, Department of Microsystems Engineering - IMTEK, University of Freiburg, Georges-Köhler-Allee 102, 79110 Freiburg, Germany
}%
\author{Ingo Breunig}
\author{Karsten Buse}
\affiliation{ 
	Laboratory for Optical Systems, Department of Microsystems Engineering - IMTEK, University of Freiburg, Georges-Köhler-Allee 102, 79110 Freiburg, Germany
}%
\affiliation{%
	Fraunhofer Institute for Physical Measurement Techniques IPM, Heidenhofstraße 8, 79110 Freiburg, Germany
}%

\date{\today}

\begin{abstract}
Eigenfrequency tuning in microresonators is useful for a range of applications including frequency-agile optical filters and tunable optical frequency converters. In most of these applications, eigenfrequency tuning is achieved by thermal or mechanical means, while a few non-centrosymmetric crystals such as lithium niobate allow for such tuning using the linear electro-optic effect. Potassium tantalate-niobate ($\mathrm{KTa}_{1-x}\mathrm{Nb}_{x}\mathrm{O}_{3}$ with $0\leq x \leq1$, KTN) is a particularly attractive material for electro-optic tuning purposes. It has both non-centrosymmetric and centrosymmetric phases offering outstandingly large linear as well as quadratic electro-optic coefficients near the phase transition temperature. We demonstrate whispering-gallery resonators (WGRs) made of KTN with quality factors of $Q>10^{7}$ and electro-optic eigenfrequency tuning of more than 100 GHz at $\lambda=1040$~nm for moderate field strengths of $E=250$~V/mm. The tuning behavior near the phase transition temperature is analyzed by introducing a simple theoretical model. These results pave the way for applications such as electro-optically tunable microresonator-based Kerr frequency combs.   
\end{abstract}

\maketitle
\section{\label{sec:level1}Introduction}
Whispering-gallery resonators (WGRs) are spheroidically-shaped monolithic dielectrics guiding light via total internal reflection. Their high quality factors and the resulting high power enhancements plus small mode volumes of the trapped light\cite{Strekalov2016} make them an ideal platform for numerous applications ranging from molecular sensors\cite{Foreman2015} to nonlinear-optical frequency conversion devices such as optical parametric oscillators\cite{Breunig2016} and frequency combs.\cite{Gaeta2019} When light is coupled into a WGR, modes build up. Their respective eigenfrequencies are a function of the optical path length inside the resonator, which depends on the geometric size of the WGR and its effective refractive index and can be calculated as\cite{Strekalov2016} 
\begin{equation}
\nu\approx m\frac{c_{0}}{2\pi R n},
\label{Eq1}
\end{equation}
where $c_{0}$ is the vacuum speed of light, $R$ is the major radius of the resonator, $m$ describes the number of oscillations in the equatorial plane of the resonator and $n$ is the refractive index of the bulk material the microresonator is made of. Hence, it is obvious that the eigenfrequency of a mode ($m$ is constant) can be altered by changing the size $R$ or the refractive index $n$. Tuning the eigenfrequencies can lead to a number of useful applications, such as frequency-tunable optical filters\cite{Monifi2012} and tunable optical frequency converters.\cite{Werner2017}  \\
To achieve eigenfrequency tuning, one can employ thermal, mechanical and electro-optic means.\cite{Strekalov2016} Thermal tuning is the most commonly applied technique. While it can lead to very large eigenfrequency tunings of hundreds of GHz - a change in temperature of 1~mK can shift the eigenfrequencies by a full linewidth\cite{Strekalov2016} - it is also a rather slow technique, especially in larger resonators. Mechanical tuning is faster with speeds in the kHz-range, but limited in the achievable eigenfrequency tuning to tens of GHz while also requiring sophisticated resonator designs, thus making the manufacturing process more difficult.\cite{Werner2017} In principle, electro-optic tuning, which has the major advantage of being quasi-instantaneous, is also available in all material platforms.\cite{Strekalov2016} Here, the application of a static electric field $E$ changes the refractive index $n$ according to the formula\cite{Saleh2007} 
\begin{equation}
\Delta n = -\frac{1}{2}n^{3}\left(\overline{r}E+\overline{s}E^{2}+...\right),
\label{Eq2}
\end{equation}
where $\overline{r}$ is the Pockels- and $\overline{s}$ is the DC-Kerr coefficient. Higher-order contributions to Eq.\,(\ref{Eq2}) are much weaker and can thus be neglected since $\overline{s}\neq 0$ in all materials.\cite{Saleh2007} In our experiment the light polarization and the external static electric field share the same direction and are both aligned to one of the principal axes of the crystal; thus, a scalar description can be used for simplicity. If we assume $\Delta n/n \ll 1$ as well as $\Delta\nu/\nu \ll 1$, we obtain
\begin{equation}
\Delta\nu=-\nu\frac{\Delta n}{n}.
\label{Eq3}
\end{equation}
In non-centrosymmetric materials, the leading term of the refractive index change depends linearly on the applied electric field: this is known as the Pockels-effect. It was shown to be strong in standard nonlinear-optical materials such as lithium niobate (LiNbO$_{3}$) and lithium tantalate (LiTaO$_{3}$).\cite{Shih1982, Minet2019} For a standard non-centrosymmetric material such as MgO-doped z-cut congruent LiNbO$_{3}$, an electric field of $E=250$~V/mm applied along the z-axis, where $\overline{r}\approx 30$~pm/V\cite{Fries1991} and $n\approx 2.19$\cite{Umemura2014} for light with a wavelength of 633~nm, induces a refractive index change of $\Delta n\approx -3.9\times 10^{-5}$ according to Eq.\,(\ref{Eq2}). If this is plugged into Eq.\,(\ref{Eq3}), this corresponds to an eigenfrequency shift of $\Delta\nu\approx 8.5$~GHz. In centrosymmetric materials, the Pockels-effect vanishes, i.e.\,$\overline{r}=0$ in Eq.\,(\ref{Eq2}). Here, eigenfrequency tuning schemes were implemented using the generation of conduction-band electrons by laser pulses, shifting the eigenfrequencies by hundreds of GHz.\cite{Preble2007} At the same time, however, the quality factor is significantly reduced. An alternative method makes use of the AC-Kerr effect.\cite{Yoshiki2016} Here, however, a second pump laser is needed and only small frequency shifts of hundreds of MHz can be observed. The most obvious choice for electro-optic eigenfrequency tuning in this material class would be to make use of the DC-Kerr effect as described by Eq.\,(\ref{Eq2}), which is present in all materials. However, in most materials it is very weak and thus neglected.\cite{Nakamura2008} For standard centrosymmetric materials such as flint glass (Schott SF6), for the same wavelength and applied electric field as in the previous examples, one would expect changes of $\left|\Delta n\right|\approx 3.2\times 10^{-11}$ and $\left|\Delta\nu\right|\approx 8$~kHz,\cite{Weber2002} orders of magnitude below those of LiNbO$_{3}$. Even glasses with much higher DC-Kerr coefficients such as As$_{2}$S$_{3}$ would only give $\left|\Delta n\right|\approx 3.4\times 10^{-9}$ and $\left|\Delta\nu\right|\approx 660$~kHz.\cite{Weber2002}\\
Let us now turn our attention to potassium tantalate-niobate (KTa$_{1-x}$Nb$_{x}$O$_{3}$ with $0\leq x \leq 1$, KTN). KTN undergoes a first-order phase transition from a crystallographic non-centrosymmetric tetragonal ferroelectric (point group 4mm) to a centrosymmetric cubic paraelectric (point group m3m) state\cite{Gupta2006} at a temperature $T_{0}$ depending linearly on the KNbO$_{3}$ fraction $x$.\cite{Triebwasser1959} In the non-centrosymmetric phase, i.e.\,for temperatures $T<T_{0}$, $\overline{r} \neq 0$ in Eq.\,(\ref{Eq2}) makes the Pockels-effect the most significant contribution to eigenfrequency tuning and higher-order effects neglectable. Close to the phase-transition temperature $T_{0}$, KTN shows outstandingly high linear electro-optic coefficients of $\overline{r} \approx 3000$~pm/V.\cite{Loheide1993} Taking into account its refractive index of $n=2.29$ at 633~nm,\cite{Newnham2005} the same applied external electric field of $E=250$~V/mm would thus lead to a refractive index change of $\Delta n\approx 4.5\times 10^{-3}$ and a corresponding eigenfrequency shift of $\Delta\nu\approx 930$~GHz. These values are two orders of magnitude larger than those of LiNbO$_{3}$. When KTN is heated to temperatures $T>T_{0}$, it transfers to its centrosymmetric state, making $\overline{r}=0$ and thus the DC-Kerr effect the strongest contribution in Eq.\,(\ref{Eq2}), with $\overline{s}=2.9\times10^{-15}~\mathrm{m}^{2}/\mathrm{V}^{2}$ close to $T_{0}$.\cite{Newnham2005} Again looking at a wavelength of 633~nm and an external electric field of 250~V/mm, this would lead to a refractive index change $\left|\Delta n\right|\approx 1.1\times 10^{-3}$ and an eigenfrequency change $\left|\Delta\nu\right|\approx 225$~GHz, even outperforming the linear electro-optic effect in LiNbO$_{3}$ by almost two orders of magnitude and the values for typical glasses by six and more orders. The reason behind the very high DC-Kerr coefficients of KTN lies in its first-order phase transition from a tetragonal 4mm to a cubic m3m state.\cite{Gupta2006} Another material undergoing the same type of phase transition is BaTiO$_{3}$,\cite{Gupta2006} which gives values of $\left|\Delta n\right|\approx 10^{-3}$ and $\left|\Delta\nu\right|\approx 200$~GHz,\cite{Newnham2005} comparable to those of KTN. The outstanding electro-optic coefficients of KTN and its wide transparency range from 390 to 5000~nm,\cite{Geusic1964} have led to KTN being discovered for a number of applications including laser scanners,\cite{Roemer2014} lenses with variable focal lengths\cite{Imai2012} and thin-film waveguides.\cite{Jia2018} While we have demonstrated first results obtained with KTN microresonators earlier,\cite{Szabados2018} in this contribution, we expand these findings by showing an in-depth analysis of the influence of the phase transition on the eigenfrequency tuning introducing a simple theoretical model. 
\section{Experimental procedure}
\begin{figure*}
\includegraphics{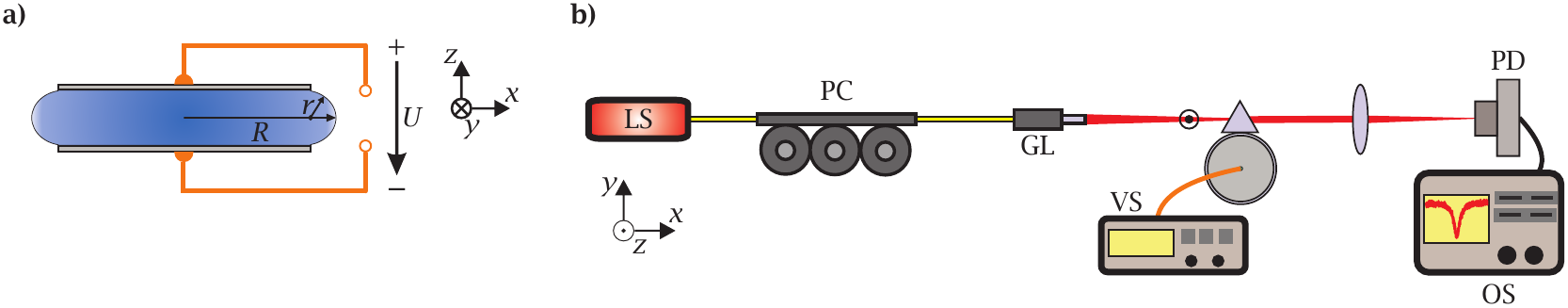}
\caption{\label{Fig1}\textbf{a)} Side-view of the resonator. Its geometry is characterized by its major radius $R=1$~mm and its minor radius $r=0.4$~mm. The resonator has chromium electrodes attached to its top and bottom sides being connected to a voltage source. \textbf{b)} Experimental setup for microresonator characterization and electro-optic eigenfrequency tuning measurements. A fiber-coupled laser source (LS) with the fiber passing through a polarization controller (PC) emits light at a wavelength of 1040~nm. When focused onto a rutile prism through a gradient-index lens (GL), which is in close proximity to the KTN resonator, light can be coupled into the resonator. The resulting transmission spectrum can be monitored using a photodetector (PD) connected to an oscilloscope (OS). Additionally, the electrodes of the resonator are connected to a voltage source (VS).}
\end{figure*}

To fabricate a microresonator, we start with a piece of z-cut KTN of $10\times10~\mathrm{mm}^{2}$ and 1.2~mm thickness. The composition used is KTa$_{0.57}$Nb$_{0.43}$O$_{3}$ corresponding to a phase transition temperature of approximately $T_{0}=52~^{\circ}\mathrm{C}$.\cite{Triebwasser1959} Since most of the manufacturing process takes place at room temperature and since higher temperatures are carefully avoided, the KTN is in its non-centrosymmetric, ferroelectric state. To allow for the application of electric fields, chromium electrodes are evaporated on the +$z$- and -$z$-sides of the crystal. Subsequently, a femtosecond laser source emitting at a wavelength of 388~nm with a repetition rate of 2~kHz and 300~mW average output power is employed to cut out a cylindrical preform of the crystal. Then, the KTN cylinder is soldered onto a metal post for easier handling. Again using the femtosecond laser source, we shape a resonator with a geometry as displayed in Fig.\,\ref{Fig1}a) with a major radius of $R=1$~mm and a minor radius of $r=0.4$~mm. Afterwards, to achieve optical-grade surface quality, we polish the rim with a diamond slurry.\\
After the resonator is prepared, it is transferred to an optical setup as shown in Fig.\,\ref{Fig1}b). Here, the resonator is placed on a mount that is temperature-controlled and -stabilized to $\pm5$~mK. The laser we used for the experiments has a center wavelength of 1040~nm and can be tuned across tens of GHz while maintaining a linewidth in the kHz-range. The output power is set to be approximately 1~mW. It is fiber-coupled with the fiber passing polarization controllers to be able to choose the light polarization freely. Throughout the experiments, it is chosen to be parallel to the rotational axis of the resonator as shown in Fig.\,\ref{Fig1}b). Then, the light passes a gradient-index lens which focuses the light on a rutile prism placed on the same mount as the resonator. When the prism is close enough to the resonator, light can be coupled to the latter when the incoming light matches an eigenfrequency as described in Eq.\,(\ref{Eq1}). To be able to apply electric fields to the resonator, a voltage source is connected to the electrodes. Finally, the light is focused on a photodetector connected to an oscillocope to allow us for the monitoring of the transmission spectrum of the resonator. By monitoring the frequency shift of the laser light and the transmission spectrum shift of the resonator, the eigenfrequency tuning can be determined.   
\section{Results and discussion}
\begin{figure*}
	\includegraphics{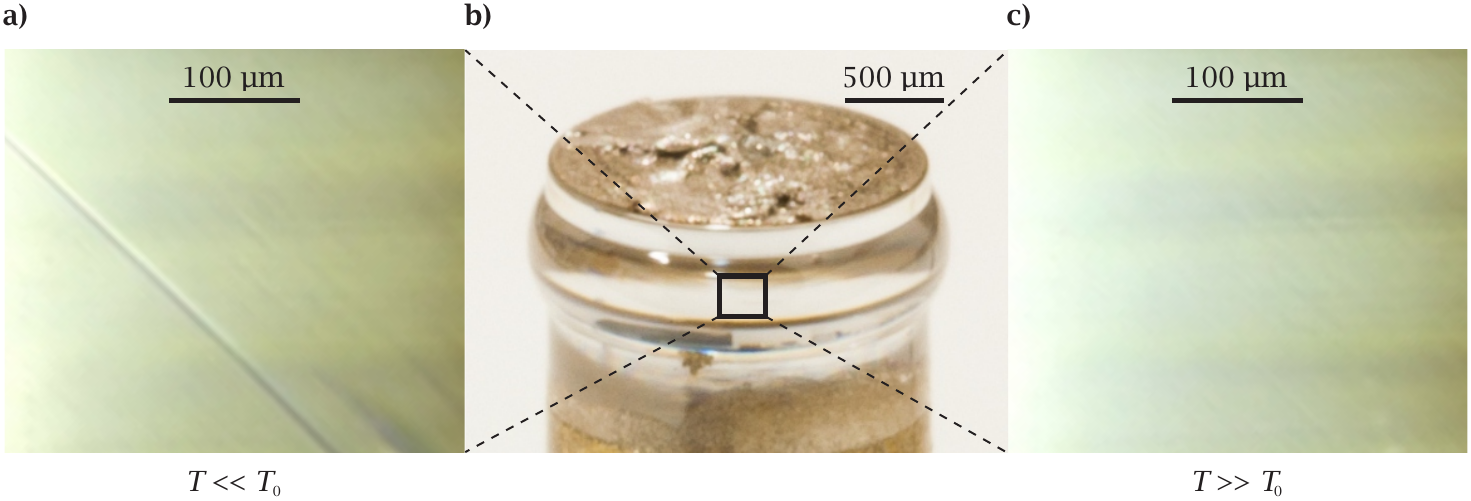}
	\caption{\label{Fig2}\textbf{a)} Close-up of the whispering gallery resonator rim as shown in b). For $T\ll T_{0}$, KTN is in its ferroelectric phase. Here, domain walls can be observed, which lead to scattering losses.  \textbf{b)} KTN microresonator with a diameter of 2~mm and chromium electrodes on the top and bottom sides. For easier handling, the resonator is attached to a brass post. \textbf{c)} For $T\gg T_\mathrm{0}$, no ferroelectric domain walls are observed since KTN is in its paraelectric phase.   }
\end{figure*}

\begin{figure*}
	\includegraphics{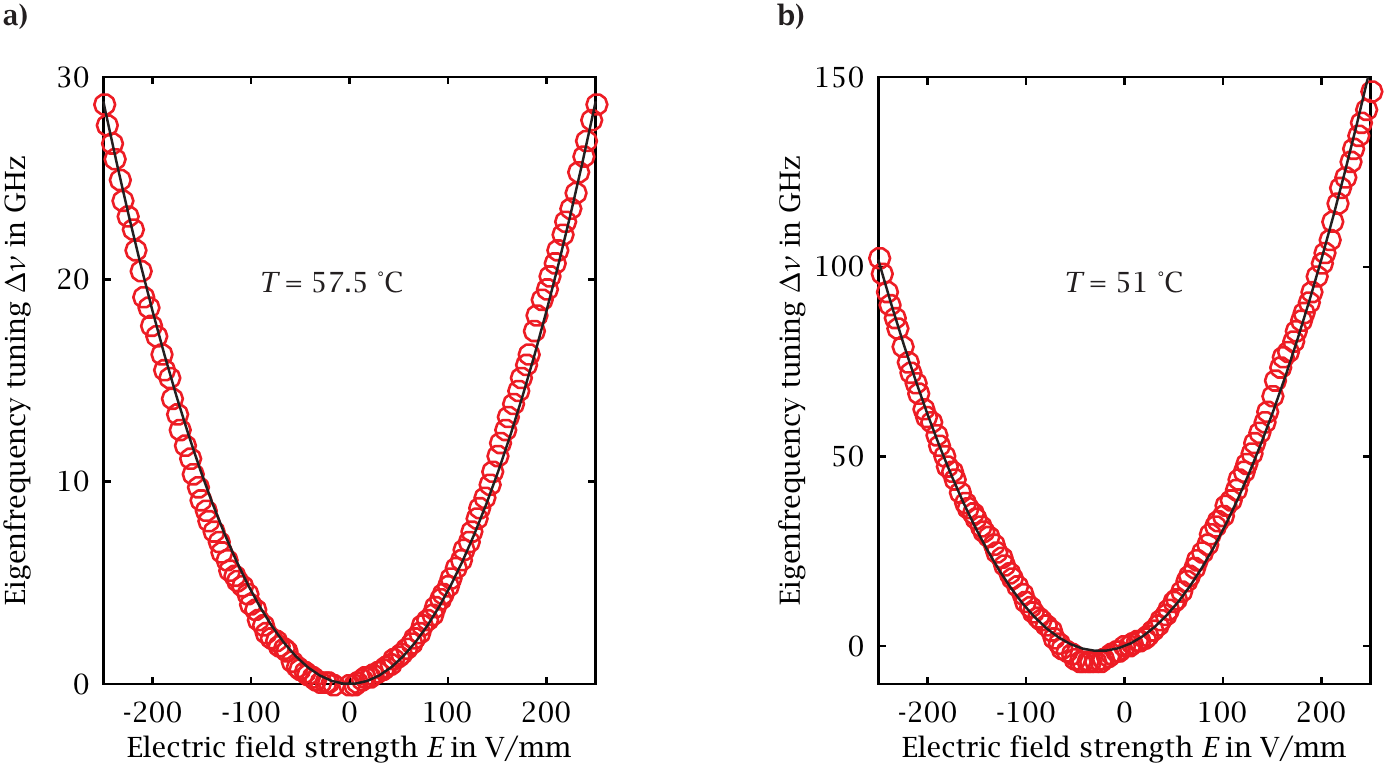}
	\caption{\label{Fig3} The red symbols describe the experimental data, while the solid black curves display the fit to the data using Eq.\,(\ref{Eq4}). \textbf{a)} Several degrees above the phase transition temperature of $T_{0}=52$~$^{\circ}\mathrm{C}$, the eigenfrequency tuning behavior is quadratic. \textbf{b)} Near the phase transition temperature, one can observe a mixture of linear and quadratic eigenfrequency tuning behavior. }
\end{figure*}

\begin{figure*}
	\includegraphics{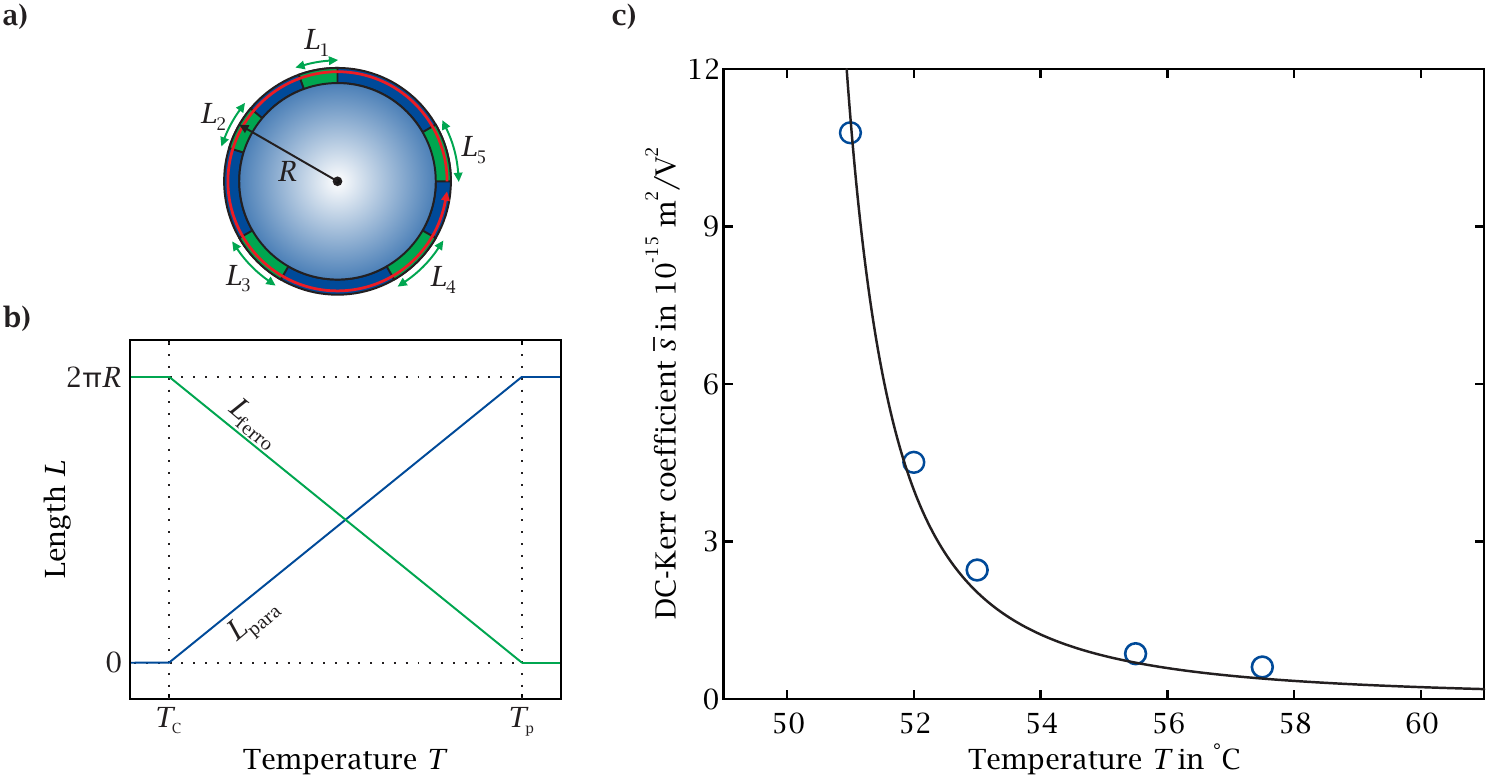}
	\caption{\label{Fig4}\textbf{a)} Top view of a microresonator with radius $R$. The red arrow indicates the laser light traveling around the rim. The dark blue parts are in the para-, green parts in the ferroelectric phase.  \textbf{b)} For $T < T_\mathrm{C}$, the entire resonator is in the ferroelectric phase of KTN as visualized by the green line. For $T>T_\mathrm{p}$, on the contrary, there are no ferroelectric parts anymore, thus $L_\mathrm{ferro}=0$. Between these two temperatures, i.e.\, for $T_\mathrm{C}<T<T_\mathrm{p}$, we expect the length of the ferroelectric regions the light experiences when traveling around the resonator rim to decrease linearly; accordingly, the length of the paraelectric regions $L_\mathrm{para}$ increases linearly (blue line). \textbf{c)} The DC-Kerr coefficient $\overline{s}$ (blue symbols), with values determined from fits as the ones presented in Fig.\,\ref{Fig3}, is shown to follow the Curie-Weiss law (black, solid line) as described by Eq.\,(\ref{Eq5}). }
\end{figure*}

In a first step, the resonator was kept at room temperature and thus in its ferroelectric, non-centrosymmetric state. In this state, it was impossible to couple light into it and thus no eigenfrequency tuning measurements could be carried out. The reason for this is revealed by taking a closer look at the microresonator rim using a microscope: a typical result is shown in Fig.\,\ref{Fig2}a). In the ferroelectric phase, the spontaneous polarization of the material leads to the formation of parallel $90^{\circ}$ domain walls known from literature for KTN.\cite{Gupta2006} These domain walls induce scattering, which in our case is such severe that they prevent the build-up of modes.  
Subsequently, we increased the temperature to $T\gg T_{0}$, where KTN is in its paraelectric, centrosymmetric phase. Since there is no spontaneous polarization in this phase, obviously there can be no ferroelectric domain walls leading to losses, as a close-up taken with a microscope also clearly displays (Fig.\,\ref{Fig2}c)). For these temperatures, whispering-gallery modes can form with intrinsic quality factors of up to $Q=1.3\times10^{7}$. Increasing the laser frequency over a wider range allows us for the determination of the free spectral range (FSR) of the resonator, which is 21~GHz. When external static electric fields of up to $E=\pm 250$~V/mm are applied, we observe a quadratic eigenfrequency tuning behavior (Fig.\,\ref{Fig3}a)) with maximum eigenfrequency tunings of approximately 30~GHz. The quadratic behavior is expected from Eqs.\,(\ref{Eq2}) and (\ref{Eq3}), since in this phase $\overline{r}=0$. 
As the highest DC-Kerr coefficients $\overline{s}$ are expected near the phase-transition temperature $T_{0}$,\cite{Nakamura2008} we subsequently decreased the temperature and conducted the same eigenfrequency tuning measurements for a number of different temperatures. At temperatures down to $55.5~^{\circ}\mathrm{C}$, we see purely quadratic behavior. Thus, down to here, it appears to be $\overline{r}=0$ in Eq.\,(\ref{Eq2}). Below these temperatures, however, the tuning curves start looking slightly differently as shown in Fig.\,\ref{Fig3}b) for $T=51~^{\circ}\mathrm{C}<T_{0}$. The maximum eigenfrequency tuning achieved is 150~GHz for $E=250$~V/mm, while for $E=-250$~V/mm this value is approximately 100~GHz. Since the asymmetric behavior starts becoming obvious at temperatures in close proximity to the phase transition temperature $T_{0}$ and since it is known that the composition of the KTN crystals may exhibit spatial inhomogeneities\cite{Fujiura2005} leading to locally different phase transition temperatures,\cite{Triebwasser1959} we introduce a simple model to explain this behavior. As we observe purely quadratic eigenfrequency tuning down to $T=55.5~^{\circ}\mathrm{C}$, we assume this temperature to be the minimum temperature at which the entire crystal is in its paraelectric, centrosymmetric state, i.e.\,$T_{\mathrm{p}}=55.5~^{\circ}\mathrm{C}$. If the temperature is decreased further, we observe that for $T<49.5~^{\circ}\mathrm{C}$ no modes can be identified anymore. Thus, we assume $T_\mathrm{C}=49.5~^{\circ}\mathrm{C}$. Below this temperature, the entire crystal is in its ferroelectric, non-centrosymmetric state. Between these temperatures, we assume the volume fraction of the ferroelectric regions to decrease linearly. Introducing $L_{i}$ as the length of the $i$th ferroelectric domain around the circumference of the microresonator, with growing temperatures for $T_\mathrm{C}<T<T_{p}$ we obtain $0<L_\mathrm{ferro}(T)=\sum_{i=1}^{N}L_{i}(T)<2\pi R$ as visualized in Fig.\,\ref{Fig4}b). The paraelectric regions show the opposite behavior since $L_\mathrm{para}(T)+L_\mathrm{ferro}(T)=2\pi R$, where $R$ is the major radius of the microresonator. Since the first-order electro-optic effect is much stronger than the second-order one, we neglect the quadratic contribution (Eq.\,(\ref{Eq2})) in the ferroelectric regions. Thus, we end up with an eigenfrequency shift formula of 
\begin{equation}
\Delta\nu = \frac{1}{2}\nu n^{2}\left[\frac{L_\mathrm{ferro}(T)}{2\pi R} \overline{r} E+\frac{L_\mathrm{para}(T)}{2\pi R} \overline{s} E^{2}\right]
\label{Eq4}
\end{equation}
when Eq.\,(\ref{Eq2}) is inserted in Eq.\,(\ref{Eq3}). 
When fitting Eq.\,(\ref{Eq4}) to our data in Fig.\,\ref{Fig3}b), one can see that a mix of first- and second-order nonlinearities describes the eigenfrequency tuning behavior very well. The values obtained for the DC-Kerr coefficient $\overline{s}$ are displayed in Fig.\,\ref{Fig4}c). For ferroelectric materials undergoing a first-order phase change such as KTN, the DC-Kerr coefficients of KTN are expected to follow the Curie-Weiss law\cite{Nakamura2008}
\begin{equation}
\overline{s}\propto \left(T-T_\mathrm{C}\right)^{-2}.
\label{Eq5}
\end{equation}
In theory, the DC-Kerr coefficient $\overline{s}$ should be divergent at $T=T_{0}$. In reality, however, this is not the case most likely due to the spatial inhomogeneities in the crystals.\cite{Fujiura2005} Thus, $T_\mathrm{C}$ is used with $T_\mathrm{C}<T_{0}$ since this is the temperature below which the entire crystal is in its non-centrosymmetric, ferroelectric state. The coefficients seem to indeed follow the Curie-Weiss law as shown by the fit in Fig.\,\ref{Fig4}c). The maximum DC-Kerr coefficient was determined to be $\overline{s}=1.08\times10^{-14}~\mathrm{m}^{2}/\mathrm{V}^{2}$. While this is approximately a factor of three higher than some previously published values for the material of approximately $3\times10^{-15}~\mathrm{m}^{2}/\mathrm{V}^{2}$,\cite{Newnham2005, Fujiura2005} also even higher values of $2.2\times10^{-14}~\mathrm{m}^{2}/\mathrm{V}^{2}$ and $6.94\times10^{-14}~\mathrm{m}^{2}/\mathrm{V}^{2}$ can be found in literature exceeding our determined value by more than a factor of six.\cite{Imai2007, Chang2013}    
Thus, while our determined value depends on a model shown in Fig.\,\ref{Fig4}b), it is within the value range one finds in literature.
\section{Outlook}
The resonator used in this work was manufactured far below the phase transition temperature $T_{0}$ in the ferroelectric phase, where domain walls are visible (Fig.\,\ref{Fig2}a)). The measurements, however, were carried out in the paraelectric phase and in close proximity of $T_{0}$. Thus, the resonator undergoes a phase transition after the manufacturing process. The determined quality factor $Q=1.3\times10^{7}$ contains losses due to surface scattering and material absorption.\cite{Strekalov2016} While the latter cannot be altered for a given material, surface scattering might not have been reduced to a minimum in this contribution, potentially leaving room for further improvement. While we cannot comment on the influence of a phase transition on the surface quality, manufacturing the resonator in its centrosymmetric phase would certainly ease its inspection since there are no ferroelectric domain walls in this phase (Fig.\,\ref{Fig2}c)). One obvious way to do this would be to heat the resonators constantly to temperatures $T>T_\mathrm{p}$. However, for the composition used in this contribution, this is highly impractical. A more elegant approach would be to use a different composition of KTN with a phase transition temperature a few degrees below room temperature. This way, one would not complicate the manufacturing process, while keeping the high DC-Kerr coefficients within easy reach. Also, heating to higher temperatures would become unnecessary.\\
KTN microresonators might also be a potential platform for Kerr frequency combs. Tunability for frequency combs is greatly beneficial for applications such as optical frequency synthesis\cite{Cundiff2001} and wavelength-division-multiplexed coherent communications.\cite{Pfeifle2014} To achieve this, mechanical actuation\cite{Papp2013} can be used. Also, linear electro-optic tuning has been implemented in an aluminum nitride microresonator.\cite{Jung2014} These methods, however, provide only small tuning ranges compared to a typical free spectral range. Larger tuning can be achieved by heating or cooling a microresonator;\cite{Xue2015} this has the drawback of being rather slow. Since we demonstrated tuning over more than an FSR in KTN microresonators, if Kerr combs could be realized on this platform, they would come with a fast and strong tuning knob. 
\section{Summary}
In this contribution, we have demonstrated electro-optic eigenfrequency tuning in a microresonator made of potassium tantalate-niobate (KTN). With the KTN entirely in its ferroelectric phase ($T\leq 49.5~^{\circ}\mathrm{C}$), no light can be coupled into the resonator. When it is heated to temperatures surpassing the phase-transition temperature $T_{0}=52~^{\circ}\mathrm{C}$ by a few degrees ($T\geq55.5~^{\circ}\mathrm{C}$), so that it is fully in its paraelectric phase, however, whispering-gallery modes build up with quality factors of up to $Q=1.3\times10^{7}$. For static external electric fields $E$, quadratic electro-optic tuning is shown for temperatures $T\geq 55.5~^{\circ}\mathrm{C}$ while for temperatures $49.5<T<55.5~^{\circ}\mathrm{C}$, a mixture of first- and second-order electro-optic eigenfrequency tuning contributions is observed. This is attributed to spatial compositional inhomogeneities in the KTN crystal, leading to locally different phase transition temperatures. The DC-Kerr coefficients are shown to follow the Curie-Weiss law for a ferroelectric material undergoing a first-order phase change. The maximum DC-Kerr coefficient is determined to be $\overline{s}=1.08\times10^{-14}~\mathrm{m}^{2}/\mathrm{V}^{2}$ at $51~^{\circ}\mathrm{C}$. The highest measured value for the eigenfrequency tuning is 150~GHz at $E=250$~V/mm. These results may be considered a first step towards unveiling the full potential of KTN microresonators for sophisticated applications such as electro-optically tunable Kerr frequency combs.
\begin{acknowledgments}
	The authors thank D. Rutsch (Fraunhofer IPM) for technical support.
\end{acknowledgments}
\nocite{*}
\bibliography{ktnbib}

\end{document}